\documentclass[a4paper, pra, twocolumn, superscriptaddress]{revtex4-1}
\usepackage{amsthm, amsmath, amssymb, amsfonts}
\usepackage{physics}
\usepackage{graphicx}
\usepackage{lipsum}
\usepackage{quantikz}
\usepackage{hyperref}
\usepackage{graphicx}
\usepackage{floatrow}
\usepackage{graphicx}
\usepackage{ragged2e} 
\usepackage[label font=bf]{subfig}
\usepackage{caption}
\newtheorem{theorem}{Theorem}

\begin{document}

\newcommand{\uv}{Departament de Física Teòrica and IFIC, Universitat de València-CSIC, 46100 Burjassot (València), Spain}

\title{Efficient protocol to estimate the Quantum Fisher Information Matrix for Commuting-Block Circuits}

\author{Rafael Gómez-Lurbe}
\affiliation{\uv}

\begin{abstract}
The Quantum Fisher Information Matrix (QFIM) is a fundamental quantity in various subfields of quantum physics. It plays a crucial role in the study of parameterized quantum states, as it quantifies their sensitivity to variations in its parameters. Recently, the QFIM has been successfully employed to enhance the optimization of variational quantum algorithms. However, its practical applicability is often hindered by the high resource requirements for its estimation. In this work, we introduce a novel protocol for computing the off-block-diagonal elements of the QFIM between different layers in a particular class of variational quantum circuits, known as \emph{commuting-block} circuits. Our approach significantly reduces the quantum resources required, specifically lowering the number of distinct quantum state preparations from \(O(m^2)\) to \(O(L^2)\), where \(m\) is the total number of parameters and \(L\) is the number of layers in the circuit. Consequently, our protocol also minimizes the number of classical measurements and post-processing operations needed to estimate the QFIM, leading to a substantial improvement in computational efficiency.
\end{abstract}

\maketitle

\section{Introduction}

Quantum computing represents a transformative shift in computational science, offering the ability to solve problems beyond the reach of classical algorithms by harnessing key principles of quantum mechanics, such as superposition, entanglement, quantum interference and quantum contextuality ~\cite{nielsen2010quantum}.

Despite promising theoretical results, implementing established quantum algorithms on current devices remains a significant challenge due to hardware limitations and error prevalence. At present, quantum processors operate in the so-called Noisy Intermediate-Scale Quantum (NISQ) era~\cite{preskill2018nisq}, characterized by a moderate number of qubits (on the order of hundreds) that are susceptible to various noise sources. These limitations severely constrain the depth and accuracy of quantum computations, making it difficult to achieve the fault-tolerance required to execute the most prominent quantum algorithms that offer a theoretical advantage over their classical counterparts.

In this context, Variational Quantum Algorithms (VQAs) have emerged as a promising strategy to leverage the computational capabilities of NISQ devices ~\cite{Cerezo_2021}. VQAs are hybrid quantum-classical algorithms in which a quantum computer prepares parameterized quantum states, while classical optimization routines adjust the parameters based on measurements performed on the quantum hardware. This hybrid approach helps mitigate some of the limitations of current quantum devices, making VQAs a practical tool for exploring quantum advantage in applications such as optimization, machine learning, and quantum simulation.

While VQAs appear to be a promising option in the current quantum computing landscape, they face several challenges. One such issue is the \textit{Barren Plateau} phenomenon~\cite{mcclean2018barren,qi2023barren,larocca2024reviewbarrenplateausvariational}, where the optimization landscape becomes exponentially flat and featureless as the number of parameters increases, making it difficult to effectively optimize the model. Another challenge is the number of quantum state preparations required to estimate the gradient of the function to be minimized, which scales with the number of parameters. A common approach in the community is the parameter-shift rule, where gradients are estimated by evaluating two different circuit preparations for each parameter in the model ~\cite{Wierichs_2022}. Consequently, while computing the cost function requires only a single circuit preparation, estimating the gradient requires \(2m\) different circuit preparations. Here, a circuit preparation refers to constructing a quantum circuit with a fixed configuration of gates and parameter values that prepares a specific quantum state. This circuit is then executed multiple times—referred to as \emph{shots}—to gather measurement statistics needed to estimate expectation values. This undesirable scaling is not present in classical machine learning, where methods such as backpropagation require roughly the same computational resources to compute the exact gradient as to evaluate the function itself~\cite{Rumelhart1986}.

Recently, the authors of~\cite{bowles2024backpropagationscalingparameterisedquantum} addressed the second issue by focusing on specific families of parameterized quantum circuits with the goal of achieving a favorable backpropagation-like scaling in terms of quantum state preparations. They introduced two types of architectures: \textit{commuting-generator circuits}, where all the generators of the unitaries introducing the trainable parameters commute, and \textit{commuting-block circuits}, where the circuit is divided into different blocks, referred to as layers, with all generators within a given layer commuting while maintaining specific commutation relations with those of other layers—namely, they either commute or anticommute.
For the first family, the authors derived a protocol to compute the gradient of a cost function with a single quantum state preparation, extending the same approach to higher-order derivatives and the Quantum Fisher Information Matrix (QFIM). For the second family, they devised a protocol to estimate the gradient of the cost function by preparing \(2L - 1\) circuits with \(N+1\) qubits, where \(L\) is the number of layers and \(N\) is the number of qubits in the original circuit (see~\cite{bowles2024backpropagationscalingparameterisedquantum}). This reduction in the number of required circuit preparations represents a significant advancement, particularly in cases where there are many parameters per layer, i.e., when \(m \gg L\), where \(m\) is the total number of parameters. Nevertheless, extending the protocol to the computation of the QFIM remained an open problem. In this work, we propose such an extension to achieve favorable scaling in the evaluation of the QFIM.

The Quantum Fisher Information (QFI) is a fundamental quantity when dealing with parameterized quantum states. Intuitively, it quantifies the sensitivity of a parameterized quantum state to variations in a given parameter, indicating how much the state changes when the parameter is modified. It plays a central role in quantum metrology~\cite{Liu_2019}, where it defines the ultimate precision limit for quantum parameter estimation through the quantum Cramér-Rao bound~\cite{Helstrom1969, Holevo1982}. Additionally, the QFI is employed in quantum many-body physics to detect multipartite entanglement~\cite{Hyllus_2012,Dell_Anna_2023} and phase transitions~\cite{Wang_2014,PhysRevB.96.104402}. When the quantum state depends on multiple parameters, the Quantum Fisher Information Matrix (QFIM) extends the concept of the QFI to capture the sensitivity of the state with respect to variations in different parameters simultaneously.

In recent years, it has also been applied in the context of VQAs to enhance the optimization of parameterized quantum circuits by leveraging information about the geometry of the Hilbert space, a method known as Quantum Natural Gradient (QNG)~\cite{stokes2020quantum}. Numerical studies indicate that this optimization algorithm can lead to faster convergence~\cite{stokes2020quantum}, help avoid certain local minima~\cite{wierichs2020avoiding}, and demonstrate strong resilience against random initializations~\cite{DellAnna:2025urw}. Furthermore, QNG has been shown to provide advantages even in the presence of noise~\cite{DellAnna:2025urw,koczor2022quantum}. Additionally, the QFIM plays a fundamental role in Variational Quantum Imaginary Time Evolution (VarQITE)~\cite{McArdle_2019}, a variational quantum algorithm designed to simulate imaginary time evolution on hybrid quantum computers, enabling the preparation of ground states in many-particle systems. Although both QNG and VarQITE show great promise, their performance can be limited by the substantial quantum and classical resources required to compute the QFIM. In general, estimating the QFIM involves \(O(m^2)\) distinct quantum circuit preparations, where \(m\) is the total number of parameters, and this must be repeated at each optimization step. This overhead can become prohibitive, making it essential to develop efficient methods to reduce this scaling and enhance the practical feasibility of QNG and VarQITE.

In light of the need for improved scaling, this work aims to introduce a protocol that reduces the number of quantum circuit preparations required to compute the QFIM for commuting-block circuits. This, in turn, enables the application of QNG and VarQITE using significantly fewer computational resources.

This work is organized as follows: In Sec.~\ref{QFIM}, we provide a brief overview of the Quantum Fisher Information Matrix. Sec.~\ref{CBCircuits} introduces the family of parameterized commuting-block circuits. In Sec.~\ref{Protocol}, we present a protocol for estimating the QFIM for commuting-block circuits with improved scaling, along with the corresponding quantum circuit required for its estimation. Finally, in Sec.~\ref{Conclusions}, we summarize our results and explore potential directions for future research, highlighting open problems in this research line.


\section{The Quantum Fisher Information Matrix}\label{QFIM}

As previously mentioned, the Quantum Fisher Information Matrix is a fundamental quantity in the study of parameterized quantum states. It characterizes the sensitivity of a quantum state to changes in its parameters, quantifying the extent to which the state is modified by such variations.

Calculating the QFIM for arbitrary quantum states, which are described by a density matrix, can be a complex task, as it requires diagonalizing the density matrix. For pure states, the expression simplifies, and the QFIM is defined as:

\begin{align}
\mathcal{F}_{ij}(\boldsymbol{\theta}) &= 4\ \text{Re} \Big[ 
\langle \partial_i \psi(\boldsymbol{\theta}) | \partial_j \psi(\boldsymbol{\theta}) \rangle \nonumber \\
&\quad - \langle \partial_i \psi(\boldsymbol{\theta}) | \psi(\boldsymbol{\theta}) \rangle 
\langle \psi(\boldsymbol{\theta}) | \partial_j \psi(\boldsymbol{\theta}) \rangle \Big],
\end{align}\label{eq:Fisher}
where \( |\psi(\boldsymbol{\theta})\rangle \) is a parameterized quantum state.

In this work, we focus on the QFIM for pure states prepared using a parameterized quantum circuit. As mentioned earlier, estimating the QFIM generally requires \( O(m^2) \) quantum circuit preparations, where \( m \) is the total number of parameters. Moreover, when employing the QFIM within QNG and VarQITE, it must be updated at each optimization step. Given this unfavorable scaling for arbitrary unstructured parameterized quantum circuits, we turn our attention to families of circuits that exhibit more efficient scaling for gradient computation—specifically, the \emph{commuting-block circuits} introduced in \cite{bowles2024backpropagationscalingparameterisedquantum}. In the next section, we introduce these circuits and discuss their properties.

\begin{figure*}[t]
        \centering
        \includegraphics[scale=0.40]{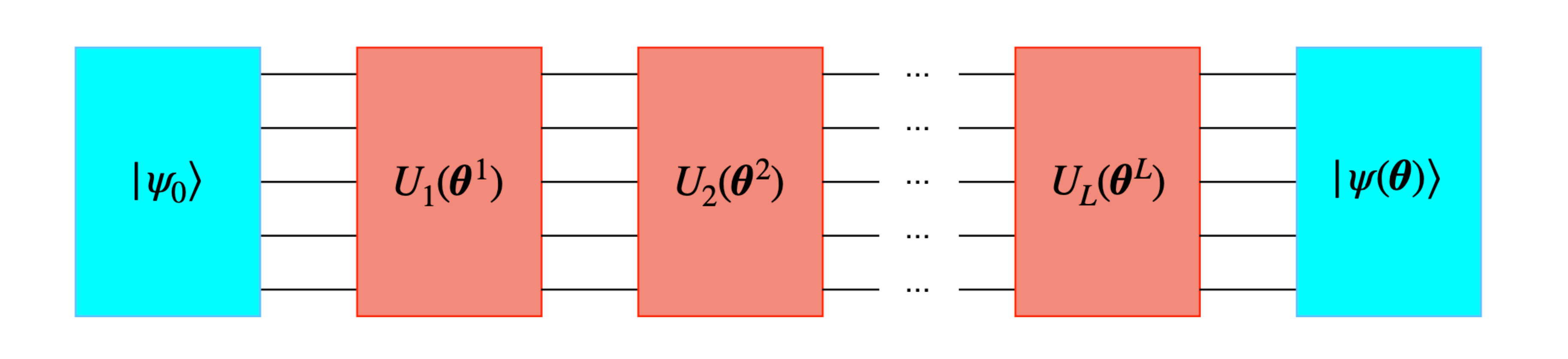}

         \caption{ \justifying General structure of a commuting-block circuit consisting of \( L \) layers. As described earlier, the generators within each layer commute, while those between different layers either commute or anticommute. The red boxes represent the individual unitary layers, the first blue box denotes the initial quantum state, and the final blue box corresponds to the parameterized quantum state obtained after applying the full quantum circuit.}
    \label{fig:General_Circuit}
  
\end{figure*}


\section{Commuting-block circuits}\label{CBCircuits}

Commuting-block circuits were introduced by the authors of \cite{bowles2024backpropagationscalingparameterisedquantum} while exploring families of parameterized quantum circuits that exhibit favorable scaling for gradient estimation.

In commuting-block circuits, the full parameterized quantum circuit is divided into distinct blocks, referred to as layers, such that the generators of all quantum operations within a given block commute. Furthermore, the generators associated with different blocks satisfy well-defined commutation relations: for each pair of layers, the generators either commute or anticommute. It is important to note that these commutation relations are not uniform across all pairs—some layers may commute with certain others while anticommute with the rest.

In the definition of commuting-block circuits, we follow the notation used in \cite{bowles2024backpropagationscalingparameterisedquantum}. More formally, consider two distinct layers, labeled 1 and 2, with sets of generators \( \mathcal{G}_1 = \{G^1_{j}\} \) and \( \mathcal{G}_2 = \{G^2_{k}\} \), respectively. By definition, all generators within each set commute among themselves. That is, all elements of \( \mathcal{G}_1 \) commute with each other, and likewise for \( \mathcal{G}_2 \). Additionally, the generators across the two layers satisfy one of the following conditions for all \( j, k \):
\begin{equation}
    [G^1_j, G^2_k] = 0 \quad \text{or} \quad \{G^1_j, G^2_k\} = 0, \quad \forall j,k.
    \label{eq:CommRel}
\end{equation}
A general parameterized quantum circuit belonging to this class and consisting of \( L \) layers is expressed as 
\begin{equation}
    U(\boldsymbol{\theta}) = \prod_j \exp(-i G^L_{j} \theta^L_{j}) \cdots \prod_k \exp(-i G^1_{k} \theta^1_{k}),
    \label{eq:Commuting-block}
\end{equation}

where the generators of each layer \( \mathcal{G}_l = \{G^l_{j}\} \) mutually commute, as defined above. Here, \( \boldsymbol{\theta} \) denotes the vector containing all the parameters across the different layers. It can be understood as the direct sum (or concatenation) of the parameter vectors associated with each layer: 
\(
\boldsymbol{\theta} = \boldsymbol{\theta}^1 \oplus \boldsymbol{\theta}^2 \oplus \dots \oplus \boldsymbol{\theta}^L.
\)
We refer to such circuits as \emph{commuting-block circuits}. The general structure of a commuting-block circuit is shown in Fig.~\ref{fig:General_Circuit}.
 Note that, while in this work we consider only parameterized gates for commuting-block circuits, as in \cite{bowles2024backpropagationscalingparameterisedquantum}, the same protocol and scaling remain valid when non-parametric gates are interleaved with parametric gates, provided they satisfy the commutation or anticommutation relations assumed in Eq. \eqref{eq:CommRel}.

While this particular construction may appear artificial and specific, a broad class of circuits falls within this definition. For example, if all operations in a quantum circuit are generated by individual Pauli strings—rather than linear combinations of them—the circuit can be classified as a commuting-block circuit. This follows from the fact that any two Pauli strings either commute or anticommute.

A specific example of a commuting-block ansatz, introduced in~\cite{bowles2024backpropagationscalingparameterisedquantum}, consists of a sequence of two alternating circuit blocks. The first block contains \(2^N/2\) gates generated by \(G_j \in \{I, Z\}^N\), restricted to operators with an odd number of \(Z\)'s. The second block comprises a single gate generated by \(X^{\otimes N}\).

They further analyze the expressivity of this particular ansatz and demonstrate that the set of unitaries realizable through sequential applications of gates generated by the aforementioned generators exceeds the maximal expressivity of general commuting-generator circuits, making it a more expressive architecture. Additionally, it is argued that non-trivial commuting-block circuits that are free from barren plateaus can be constructed. However, care must be taken to prevent circuits that are classically simulable

Thus, while a rigorous understanding of the expressivity and barren-plateau resilience of specific commuting-block architectures is still lacking, this family of circuits stands out as a promising ansatz for variational quantum algorithms.

In \cite{bowles2024backpropagationscalingparameterisedquantum}, the authors also demonstrated that for commuting-block circuits, the evaluation of gradients can be performed using a number of quantum circuit evaluations that scales with the number of layers rather than the total number of parameters. Additionally, they proposed a protocol to estimate these gradients with this favorable scaling.

Following a similar approach, we introduce a protocol to estimate the QFIM, reducing the number of required quantum circuit preparations from \( O(m^2) \) to \( O(L^2) \), where \( m \) is the total number of parameters and \( L \) is the number of layers in the circuit. In the subsequent section, we present the details of this protocol.


\section{Protocol}\label{Protocol}

The goal of this protocol is to achieve better scaling for estimating the QFIM in commuting-block circuits. Recall the definition of the QFIM:  
\begin{align}
    \mathcal{F}^{l_1 l_2}_{ij}(\boldsymbol{\theta}) &= 4\ \text{Re} \Big[ 
    \langle \partial^{l_1}_i \psi(\boldsymbol{\theta}) | \partial^{l_2}_j \psi(\boldsymbol{\theta}) \rangle \nonumber \\
    &\quad - \langle \partial^{l_1}_i \psi(\boldsymbol{\theta}) | \psi(\boldsymbol{\theta}) \rangle 
    \langle \psi(\boldsymbol{\theta}) | \partial^{l_2}_j \psi(\boldsymbol{\theta}) \rangle \Big],
    \label{eq:Fisher}
\end{align}
where the superscripts \( l_1 \) and \( l_2 \) indicate that the corresponding QFIM entries are associated with parameters from layers \( l_1 \) and \( l_2 \), respectively. These superscripts are included to explicitly track which layer each parameter belongs to when differentiating. The indices \( l_1 \) and \( l_2 \) range from 1 to \( L \), and \( i \) and \( j \) range from 1 to the number of parameters in their respective layers.

It is important to note that the QFIM is symmetric with respect to index permutations, including the superscripts corresponding to different layers. Therefore, it is sufficient to compute only the diagonal and either the upper or lower triangular elements to fully determine the QFIM.

When \(l_1 = l_2\), the entries of the QFIM can be efficiently estimated on a quantum computer. In fact, all terms for every \(i\) and \(j\), can be computed simultaneously using a single quantum circuit for a given layer \(l\). This approach, introduced in \cite{stokes2020quantum}, is referred to as the block-diagonal approximation of the QFIM. Thus, if the quantum circuit is composed of \(L\) different blocks, we would need \(L\) different quantum circuits to estimate the block-diagonal approximation of the QFIM.

Nevertheless, this does not hold for the off-block-diagonal terms, requiring a Hadamard test to estimate each upper(lower) off-block-diagonal term between different layers. In fact, the term primarily responsible for the unfavorable scaling in the computation of the QFIM is the first one on the right-hand side of Eq. \eqref{eq:Fisher}: 
\begin{equation}
    \text{Re} \Big[ 
    \langle \partial^{l_1}_i \psi(\boldsymbol{\theta}) | \partial^{l_2}_j \psi(\boldsymbol{\theta}) \rangle \Big].
    \label{eq:FisherTerm}
\end{equation}

Thus, our ultimate goal is to estimate these terms efficiently. For simplicity, we will omit the explicit dependence of \( | \psi(\boldsymbol{\theta}) \rangle \) on \( \boldsymbol{\theta} \) in the notation.  

To illustrate the protocol, we focus on estimating these terms for two specific layers, \( l_1 \) and \( l_2 \), assuming without loss of generality that \( l_2 > l_1 \). Since the QFIM is symmetric with respect to the permutation of its indices, including the layer indices, this assumption is sufficient to determine the entire matrix.

\begin{figure*}[t]
\centering
\sidesubfloat[]{\includegraphics[width=0.45\textwidth]{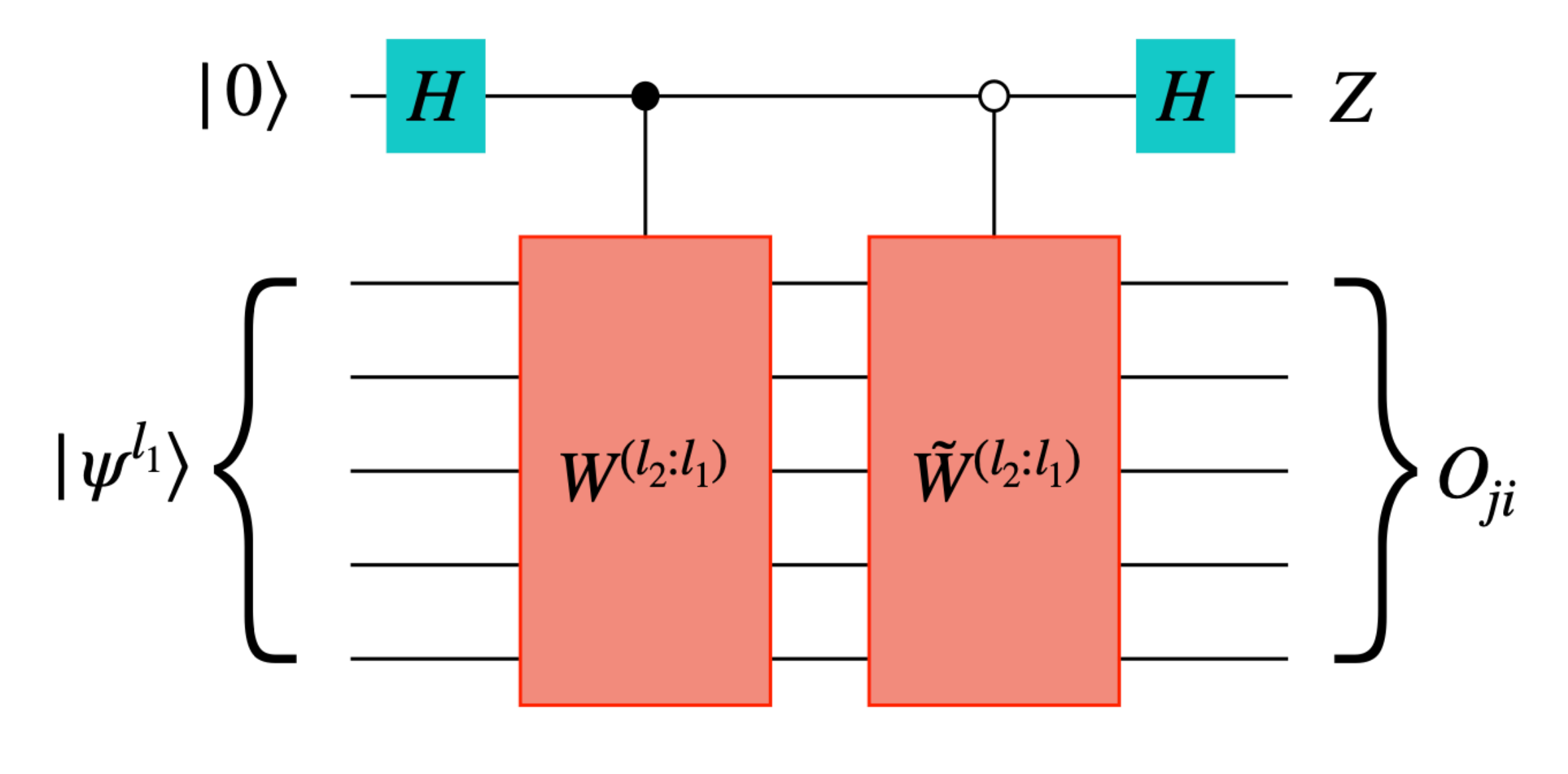}\label{fig:a}}
\hfil
\sidesubfloat[]{\includegraphics[width=0.45\textwidth]{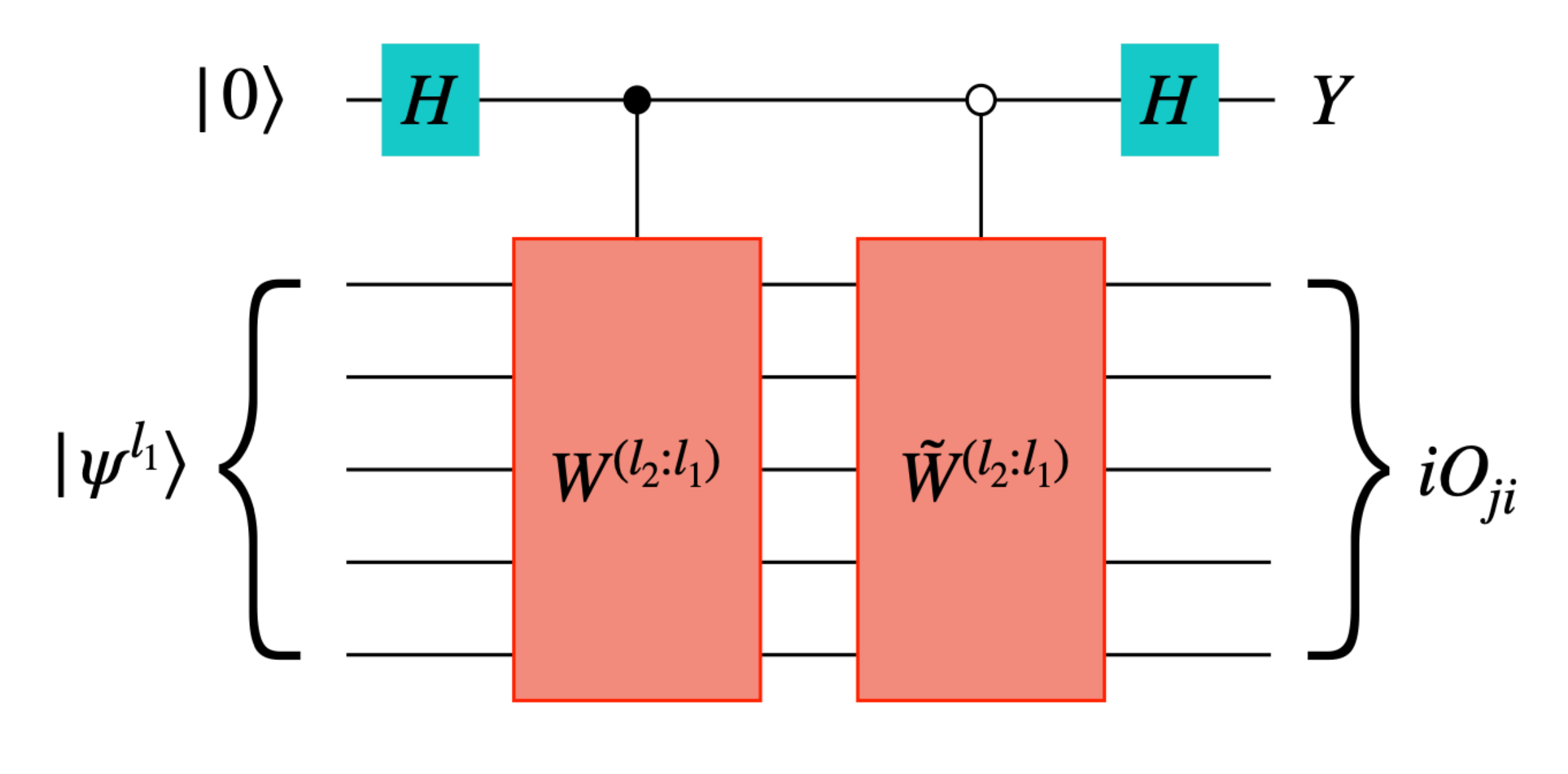}\label{fig:b}}
\caption{
\justifying Quantum circuits used to estimate the off-block-diagonal elements of the QFIM using the linear combination of unitaries method. (a) Circuit for the case where the generators commute, i.e., \([G^{l_1}_j, G^{l_2}_k] = 0\). (b) Circuit for the case where the generators anticommute, i.e., \(\{G^{l_1}_j, G^{l_2}_k\} = 0\).}
\label{fig:LinCombCircuit}
\end{figure*}

Focusing on Eq. \eqref{eq:FisherTerm}, the right-hand term is given by:
\begin{equation}
    | \partial^{l_2}_j \psi \rangle = W^{l_2} (-i G^{l_2}_j) |\psi^{l_2} \rangle,
\end{equation}
where the unitary operator \( W^{l_2} \) is defined as:
\begin{equation}
    W^{l_2} = U^L(\boldsymbol{\theta}^L) \cdots U^{l_2+1}(\boldsymbol{\theta}^{l_2+1}),
\end{equation}
with each unitary layer \(b\) given by:
\begin{equation}
    U^b(\boldsymbol{\theta}^b) = \prod_j \exp(-i G^b_{j} \theta^b_{j}).
\end{equation}
Additionally, the quantum state \( |\psi^{l_2} \rangle \), resulting from the application of the first \( l_2 \) layers of the circuit, is expressed as:
\begin{equation}
    |\psi^{l_2} \rangle = U^{l_2}(\boldsymbol{\theta}^{l_2}) \cdots U^{1}(\boldsymbol{\theta}^{1}) |\psi_{0} \rangle.
\end{equation}

Similarly, the left-hand term reads:

\begin{equation}
     \langle \partial^{l_1}_i \psi|  =  \langle \psi^{l_1} |(iG^{l_1}_i)(W^{l_1})^{\dagger}.
\end{equation}

Hence, the inner product of both terms is given by:
\begin{equation}
\begin{aligned}
    \langle \partial^{l_1}_i \psi | \partial^{l_2}_j \psi \rangle 
    &=  \langle \psi^{l_1} |(iG^{l_1}_i)(W^{l_1})^{\dagger} W^{l_2} (-i G^{l_2}_j) |\psi^{l_2} \rangle \\ 
    &= \langle \psi^{l_1} |G^{l_1}_i(W^{(l_2:l_1)})^{\dagger} G^{l_2}_jW^{(l_2:l_1)} |\psi^{l_1} \rangle,
\end{aligned}
\end{equation}

where the unitary operator \( W^{(l_2:l_1)} \) is defined as:
\begin{equation}
    W^{(l_2:l_1)} = U^{l_2}(\boldsymbol{\theta}^{l_2}) \cdots U^{l_1+1}(\boldsymbol{\theta}^{l_1+1}).
\end{equation}

Since \( G^{l_1}_i \) either commutes or anticommutes with the gates forming \( (W^{(l_2:l_1)})^{\dagger} \), we can rewrite the expression as  

\begin{equation}
    G^{l_1}_i(W^{(l_2:l_1)})^{\dagger} = (\tilde{W}^{(l_2:l_1)})^{\dagger} G^{l_1}_i.
\end{equation}

Using this identity, the inner product simplifies to  

\begin{equation}
    \langle \partial^{l_1}_i \psi | \partial^{l_2}_j \psi \rangle 
    =  \langle \psi^{l_1} |(\tilde{W}^{(l_2:l_1)})^{\dagger} G^{l_1}_i G^{l_2}_j W^{(l_2:l_1)}|\psi^{l_1} \rangle.
\end{equation}

Here, \( \tilde{W}^{(l_2:l_1)} \) represents the modified unitary incorporating the commutation or anticommutation effects of \( G^{l_1}_i \). This unitary can be constructed straightforwardly: for layers whose generators commute with \( G^{l_1}_i \), the corresponding unitaries remain unchanged; for layers whose generators anticommute with \( G^{l_1}_i \), the sign of the parameters associated with those unitaries is flipped.

Recall that we need the real part of this term. Thus, we obtain:  

\begin{equation}
    \begin{aligned}
         \text{Re} \Big[ 
    \langle \partial^{l_1}_i \psi | \partial^{l_2}_j \psi \rangle \Big] &= \frac{1}{2} \Big( 
    \langle \psi^{l_1} | (\tilde{W}^{(l_2:l_1)})^{\dagger} G^{l_1}_i  G^{l_2}_j W^{(l_2:l_1)} |\psi^{l_1} \rangle \\
    &+ \langle \psi^{l_1} | (W^{(l_2:l_1)})^{\dagger} G^{l_2}_j  G^{l_1}_i \tilde{W}^{(l_2:l_1)} |\psi^{l_1} \rangle \Big)\\
    &=\frac{1}{2} \Big( 
    \langle \psi^{l_1} | (\tilde{W}^{(l_2:l_1)})^{\dagger} G^{l_1}_i  G^{l_2}_j W^{(l_2:l_1)}\\
    &+(W^{(l_2:l_1)})^{\dagger} G^{l_2}_j  G^{l_1}_i \tilde{W}^{(l_2:l_1)} |\psi^{l_1} \rangle \Big).
    \end{aligned}
    \label{eq:RealFish}
\end{equation}

This type of term can be computed on a quantum computer using the linear combination of unitaries method \cite{ChildsWiebe2012}.  

We now consider two different scenarios:  
\begin{equation}
    [G^{l_1}_i, G^{l_2}_j] = 0 \quad \text{or} \quad \{G^{l_1}_i, G^{l_2}_j\} = 0, \quad \forall i,j.
\end{equation}  

In the case where \([G^{l_1}_i, G^{l_2}_j] = 0\), the quantum circuit required to estimate the term defined in Eq. \eqref{eq:RealFish} is shown in Fig. \ref{fig:a}. For convenience, we introduce the following operator:  
\begin{equation}
    O_{ji} \coloneq G^{l_2}_j G^{l_1}_i.
\end{equation}  

Similarly, when \(\{G^{l_1}_j, G^{l_2}_k\} = 0\), the corresponding circuit has a similar structure and is shown in Fig. \ref{fig:b}.

The proof of the protocol for both cases can be found in Appendix \ref{ProofProtocol}.

For both circuits, the observables \(O_{ji}\) required to estimate the QFIM entries between different blocks commute with each other. Consequently, for a given \(l_1\) and \(l_2\), all these observables can be simultaneously diagonalized, allowing their expectation values to be estimated using a single quantum circuit. By rotating the state into the shared eigenbasis prior to measurement, all expectation values can be obtained by sampling the circuit in the computational basis through post-processing. This approach was also utilized in \cite{bowles2024backpropagationscalingparameterisedquantum}.

As a result, a single quantum circuit is sufficient to evaluate all off-block-diagonal terms between two distinct layers, \(l_1\) and \(l_2\). Therefore, in commuting-block circuits, the number of quantum circuit preparations required to estimate the full QFIM is reduced from \(O(m^2)\) to \(O(L^2)\), where \(m\) is the total number of parameters and \(L\) is the number of layers. Since \(L < m\) always, and in some cases \(L \ll m\), this protocol can lead to a significant reduction in quantum resource requirements. Additionally, it decreases the number of measurements needed to estimate the QFIM, further improving efficiency. In particular, estimating a single expectation value with an error \(\epsilon\) requires \(O(1/\epsilon^2)\) repetitions, also known as \emph{shots}, of the quantum circuit. Consequently, reducing the number of required circuits significantly decreases the total number of shots needed during the optimization procedure. This results in a more efficient and less expensive optimization process.

The main contribution of this work is summarized on the following theorem.

\begin{theorem}
    Consider a commuting-block circuit defined by the unitary \(U(\boldsymbol{\theta})\), as given in Eq. \eqref{eq:Commuting-block}, applied to an arbitrary initial state. Then, an unbiased estimator of the QFIM for the prepared state can be obtained by classically post-processing data from \(\frac{L(L+1)}{2}\) distinct quantum circuits. Of these, \(L\) circuits correspond to the block-diagonal terms and require the same number of qubits as \(U(\boldsymbol{\theta})\). The remaining circuits, used for off-block-diagonal terms, require an additional auxiliary qubit. Furthermore, the variance of each QFIM entry estimator scales as \(\mathcal{O}(1/M)\), where \(M\) is the total number of shots used for each quantum circuit evaluation.
    \label{Theorem}
\end{theorem}

The proof of the Theorem \ref{Theorem} can be found in Appendix \ref{ProofTheorem}.


\section{Conclusions and Outlook}\label{Conclusions}

In this work, we introduced a novel protocol for computing the off-block-diagonal elements of the QFIM between different layers in a commuting-block parameterized quantum circuit. This approach significantly reduces the quantum resources required, specifically the number of distinct quantum state preparations, from \(O(m^2)\) to \(O(L^2)\), where \(m\) represents the total number of parameters and \(L\) denotes the number of layers. As a result, our protocol also decreases the number of classical measurements and the amount of post-processing needed to estimate the QFIM, leading to a substantial efficiency improvement. This advancement facilitates the application of QFIM in variational quantum algorithms, such as QNG and VarQITE.

Future research should focus on a rigorous analysis of commuting-block circuits, particularly their expressiveness as quantum machine learning (QML) models, methods for constructing them systematically, and their susceptibility to the \emph{barren plateau} phenomenon. Additionally, numerical studies should be conducted to heuristically evaluate the performance of this ansatz across various variational quantum computing tasks. These include applications such as finding the ground state of a given Hamiltonian using variational circuits, quantum machine learning problems, and quantum optimization, among others.


\section{Acknowledgments}

This work was supported by the project PID2023-152724NA-I00, with funding from MCIU/AEI/10.13039/501100011033 and FSE+, the Severo Ochoa Grant CEX2023-001292-S, Generalitat Valenciana grant CIPROM/2022/66, the Ministry of Economic Affairs and Digital Transformation of the Spanish Government through the QUANTUM ENIA project call - QUANTUM SPAIN project, and by the European Union through the Recovery, Transformation and Resilience Plan - NextGenerationEU within the framework of the Digital Spain 2026 Agenda, and by the CSIC Interdisciplinary Thematic Platform (PTI+) on Quantum Technologies (PTI-QTEP+). This project has also received funding from the European Union’s Horizon 2020 research and innovation program under grant agreement CaLIGOLA MSCA-2021-SE-01-101086123. RGL is funded by grant CIACIF/2021/136 from Generalitat Valenciana.


\bibliographystyle{unsrt}
\bibliography{references}

\onecolumngrid
\appendix
\section{Proof of Protocol}\label{ProofProtocol}

Remember that we arrived at the following equality:

\begin{equation}
    \text{Re} \Big[ 
    \langle \partial^{l_1}_i \psi | \partial^{l_2}_j \psi \rangle \Big] = \frac{1}{2} \Big( 
    \langle \psi^{l_1} | (\tilde{W}^{(l_2:l_1)})^{\dagger} G^{l_1}_i  G^{l_2}_j W^{(l_2:l_1)} + (W^{(l_2:l_1)})^{\dagger} G^{l_2}_j  G^{l_1}_i \tilde{W}^{(l_2:l_1)} |\psi^{l_1} \rangle \Big).
    \label{eq:RealFishApp}
\end{equation}

We now consider two distinct scenarios:  
\begin{equation}
    [G^{l_1}_j, G^{l_2}_k] = 0 \quad \text{or} \quad \{G^{l_1}_j, G^{l_2}_k\} = 0, \quad \forall j,k.
\end{equation}

\subsection{Commuting Layers}

We first examine the case where all generators between layers \(l_1\) and \(l_2\) commute, meaning that
\begin{equation}
    [G^{l_1}_i, G^{l_2}_j] = 0 , \quad \forall i,j.
\end{equation}

For convenience, we define the following operator:  
\begin{equation}
    O_{ji} \coloneq G^{l_2}_j G^{l_1}_i.
\end{equation}  

Since \(G^{l_2}_j\) commutes with \(G^{l_1}_i\), it follows that \(O_{ij} = O_{ji}\).

Then,

\begin{equation}
    \begin{aligned}
        \text{Re} \Big[ 
    \langle \partial^{l_1}_i \psi | \partial^{l_2}_j \psi \rangle \Big] &= \frac{1}{2} \Big( 
    \langle \psi^{l_1} | (\tilde{W}^{(l_2:l_1)})^{\dagger} O_{ij} W^{(l_2:l_1)} + (W^{(l_2:l_1)})^{\dagger} O_{ji} \tilde{W}^{(l_2:l_1)} |\psi^{l_1} \rangle \Big) \\
    &= \frac{1}{2} \Big( 
    \langle \psi^{l_1} | (\tilde{W}^{(l_2:l_1)})^{\dagger} O_{ji} W^{(l_2:l_1)}   + (W^{(l_2:l_1)})^{\dagger} O_{ji} \tilde{W}^{(l_2:l_1)} |\psi^{l_1} \rangle \Big) \\
    &=\frac{1}{4}\Bigg(
    \langle \psi^{l_1} |\Big((\tilde{W}^{(l_2:l_1)})^{\dagger}+(W^{(l_2:l_1)})^{\dagger}\Big)O_{ji}\Big(W^{(l_2:l_1)}+\tilde{W}^{(l_2:l_1)}\Big)|\psi^{l_1} \rangle \\
    &+
    \langle \psi^{l_1} |\Big((\tilde{W}^{(l_2:l_1)})^{\dagger}-(W^{(l_2:l_1)})^{\dagger}\Big)O_{ji}\Big(W^{(l_2:l_1)}-\tilde{W}^{(l_2:l_1)}\Big)|\psi^{l_1} \rangle\Bigg) \\
     &=\frac{1}{4}\Big[\langle \psi^{l_1} |(L^{+}_W)^\dagger O_{ij}L^{+}_W|\psi^{l_1} \rangle -  \langle \psi^{l_1} |(L^{-}_W)^\dagger O_{ij}L^{-}_W|\psi^{l_1} \rangle\Big] \\
    &=\frac{1}{4}\Bigg(\langle O_{ji}\rangle_{L^{+}_W|\psi^{l_1} \rangle}- \langle O_{ji} \rangle_{L^{-}_W|\psi^{l_1} \rangle}\Bigg),
    \end{aligned}
    \label{eq:AppTermC}
\end{equation}
where \(L^{\pm}\coloneq  W^{(l_2:l_1)}\pm\tilde{W}^{(l_2:l_1)}\).

As previously discussed, Eq.~\ref{eq:AppTermC} can be estimated on a quantum computer using an auxiliary qubit, by executing the circuit shown in Fig.~\ref{fig:a}.

In this circuit, the prepared state is given by:
\begin{equation}
    |\phi\rangle=\frac{1}{\sqrt{2}}\Big( |0\rangle L^{+}_W|\psi^{l_1}\rangle + |1\rangle L^{-}_W|\psi^{l_1}\rangle\Big).
\end{equation}

Measuring the expectation value of \(Z \otimes O_{ji}\) with respect to \(|\phi\rangle\) yields:

\begin{equation}
    \begin{aligned}
        \langle\phi|(Z\otimes O_{ji})|\phi\rangle &=\frac{1}{4} \Big[\langle 0 |Z|0\rangle \langle \psi^{l_1} |(L^{+}_W)^\dagger O_{ji}L^{+}_W|\psi^{l_1} \rangle \quad + \langle 1 |Z|1\rangle \langle \psi^{l_1} |(L^{-}_W)^\dagger O_{ji}L^{-}_W|\psi^{l_1} \rangle \Big] \\
        &=\frac{1}{4} \Big[\langle \psi^{l_1} |(L^{+}_W)^\dagger O_{ji}L^{+}_W|\psi^{l_1} \rangle -  \langle \psi^{l_1} |(L^{-}_W)^\dagger O_{ji}L^{-}_W|\psi^{l_1} \rangle \Big] \\
        &=\frac{1}{4} \Big[\langle O_{ji}\rangle_{L^{+}_W|\psi^{l_1} \rangle} - \langle O_{ji} \rangle_{L^{-}_W|\psi^{l_1} \rangle} \Big] \\
        &=\text{Re} \Big[ 
    \langle \partial^{l_1}_i \psi | \partial^{l_2}_j \psi \rangle \Big].
    \end{aligned}
\end{equation}

Additionally, we can also show that since all the generators for a given \(l_1\) and \(l_2\) commute, \([O_{ij},O_{km}]=0 \quad \forall i,j,k,l\):
\begin{equation}
    [O_{ji},O_{km}]=[G^{l_2}_j G^{l_1}_i,G^{l_2}_k G^{l_1}_m]=G^{l_2}_j G^{l_1}_iG^{l_2}_k G^{l_1}_m-G^{l_2}_k G^{l_1}_mG^{l_2}_j G^{l_1}_i=0.
\end{equation}

Therefore, for a given \(l_1\) and \(l_2\), all these observables can be simultaneously diagonalized, allowing their expectation values to be estimated using a single quantum circuit.

\subsection{Anticommuting layers}

Now, we consider the case where all generators between layers \(l_1\) and \(l_2\) anticommute, meaning that

\begin{equation}
    \{G^{l_1}_i, G^{l_2}_j\} = 0, \quad \forall i,j.
\end{equation}

Notice that, in this case, since \(G^{l_1}_i\) anticommutes with \(G^{l_2}_j\), it follows that \(O_{ij} = -O_{ji}\).

Thus,

\begin{equation}
    \begin{aligned}
        \text{Re} \Big[ 
    \langle \partial^{l_1}_i \psi | \partial^{l_2}_j \psi \rangle \Big] &= \frac{1}{2} \Big( 
    \langle \psi^{l_1} | (\tilde{W}^{(l_2:l_1)})^{\dagger} O_{ij} W^{(l_2:l_1)} +  (W^{(l_2:l_1)})^{\dagger} O_{ji} \tilde{W}^{(l_2:l_1)} |\psi^{l_1} \rangle \Big) \\
    &= \frac{1}{2} \Big( 
    \langle \psi^{l_1} | -(\tilde{W}^{(l_2:l_1)})^{\dagger} O_{ji} W^{(l_2:l_1)} +  (W^{(l_2:l_1)})^{\dagger} O_{ji} \tilde{W}^{(l_2:l_1)} |\psi^{l_1} \rangle \Big) \\
    &= \frac{1}{2} \Big( 
    \langle \psi^{l_1} | (W^{(l_2:l_1)})^{\dagger} O_{ji} \tilde{W}^{(l_2:l_1)} -(\tilde{W}^{(l_2:l_1)})^{\dagger} O_{ji} W^{(l_2:l_1)} |\psi^{l_1} \rangle \Big)\\
    &=\frac{1}{4}\Bigg(
    \langle \psi^{l_1} |\Big((W^{(l_2:l_1)})^{\dagger}+(\tilde{W}^{(l_2:l_1)})^{\dagger}\Big)O_{ji}\Big(\tilde{W}^{(l_2:l_1)}-W^{(l_2:l_1)}\Big)|\psi^{l_1} \rangle \\
    &+
    \langle \psi^{l_1} |\Big((W^{(l_2:l_1)})^{\dagger}-(\tilde{W}^{(l_2:l_1)})^{\dagger}\Big)O_{ji}\Big(W^{(l_2:l_1)}+\tilde{W}^{(l_2:l_1)}\Big)|\psi^{l_1} \rangle\Bigg) \\
     &=\frac{1}{4}\Big(\langle \psi^{l_1} |(L^{+}_W)^\dagger O_{ji}L^{-}_W|\psi^{l_1} \rangle -  \langle \psi^{l_1} |(L^{-}_W)^\dagger O_{ji}L^{+}_W|\psi^{l_1} \rangle\Big),
    \end{aligned}
    \label{eq:AppTermAC}
\end{equation}
where \(L^{\pm}\coloneq  W^{(l_2:l_1)}\pm\tilde{W}^{(l_2:l_1)}\).

The term \eqref{eq:AppTermAC} can also be estimated on a quantum computer using an auxiliary qubit, by executing the circuit in Fig. \ref{fig:b}.

In this circuit, the state that is prepared is:

\begin{equation}
    |\chi\rangle=\frac{1}{\sqrt{2}}\Big( |0\rangle L^{-}_W|\psi^{l_1}\rangle + |1\rangle L^{+}_W|\psi^{l_1}\rangle\Big).
\end{equation}

The expectation value of \(Y\otimes (iO_{ji})\) measured with respect to \(|\chi\rangle\) is given by:

\begin{equation}
    \begin{aligned}
        \langle\chi|\big(Y\otimes (iO_{ji})|\chi\rangle &=\frac{1}{4} \Big[\langle 0 |Y|1\rangle \langle \psi^{l_1} |(L^{-}_W)^\dagger (iO_{ji})L^{+}_W|\psi^{l_1} \rangle  + \langle 1 |Y|0\rangle \langle \psi^{l_1} |(L^{+}_W)^\dagger (iO_{ji})L^{-}_W|\psi^{l_1} \rangle \Big]\\ &=\frac{1}{4}\Big[\langle \psi^{l_1} |(L^{+}_W)^\dagger O_{ji}L^{-}_W|\psi^{l_1} \rangle - \langle \psi^{l_1} |(L^{-}_W)^\dagger O_{ji}L^{+}_W|\psi^{l_1} \rangle\Big] \\
        &=\text{Re} \Big[ 
    \langle \partial^{l_1}_i \psi | \partial^{l_2}_j \psi \rangle \Big].
    \end{aligned}
\end{equation}

Furthermore, as in the previous case, all the observables that need to be measured in the quantum circuit commute for a given \(l_1\) and \(l_2\), as shown by the following relation:  

\begin{equation}
    \begin{aligned}
        [iO_{ji},iO_{km}]&=[O_{km},O_{ji}]=[G^{l_2}_k G^{l_1}_m,G^{l_2}_j G^{l_1}_i]=G^{l_2}_k G^{l_1}_mG^{l_2}_j G^{l_1}_i - G^{l_2}_j G^{l_1}_iG^{l_2}_k G^{l_1}_m\\
        &= G^{l_2}_j G^{l_1}_mG^{l_2}_k G^{l_1}_i-G^{l_2}_j G^{l_1}_iG^{l_2}_m G^{l_1}_k=G^{l_2}_j G^{l_1}_iG^{l_2}_m G^{l_1}_k-G^{l_2}_j G^{l_1}_iG^{l_2}_m G^{l_1}_k=0.
    \end{aligned}
\end{equation}

Therefore, for a given \(l_1\) and \(l_2\), all these observables can be simultaneously diagonalized, enabling their expectation values to be estimated using a single quantum circuit.

\section{Proof of Theorem 1}\label{ProofTheorem}

Recall that we aim to estimate the QFIM, defined as:

\begin{equation}
    \mathcal{F}^{l_1 l_2}_{ij}(\boldsymbol{\theta}) = 4\ \text{Re} \Big[ 
    \langle \partial^{l_1}_i \psi(\boldsymbol{\theta}) | \partial^{l_2}_j \psi(\boldsymbol{\theta}) \rangle     - \langle \partial^{l_1}_i \psi(\boldsymbol{\theta}) | \psi(\boldsymbol{\theta}) \rangle 
    \langle \psi(\boldsymbol{\theta}) | \partial^{l_2}_j \psi(\boldsymbol{\theta}) \rangle \Big].
\end{equation}

As pointed out in \cite{stokes2020quantum}, the second term on the right-hand side can be efficiently estimated on a quantum computer for all \(l_1\) and \(l_2\), as it corresponds to the expectation values of the generators associated with the parameters being differentiated:

\begin{equation}
    \langle \partial^{l_1}_i \psi(\boldsymbol{\theta}) | \psi(\boldsymbol{\theta}) \rangle = \langle \psi^{l_1-1} | (iG^{l_1}_i) | \psi^{l_1-1}  \rangle, \ \text{and} \quad \langle \psi(\boldsymbol{\theta}) | \partial^{l_2}_j \psi(\boldsymbol{\theta}) \rangle=\langle \psi^{l_2-1} | (-iG^{l_2}_j) | \psi^{l_2-1}  \rangle.
    \label{eq:2RHS}
\end{equation}

For a given layer, since all the generators commute with each other, they can be simultaneously diagonalized, allowing all of them to be estimated in parallel using a single quantum circuit. Additionally, the computed expectation values can be reused for QFIM entries corresponding to different layers.

Now, we focus on the first term on the right-hand side:

\begin{equation}
    \text{Re} \Big[ 
    \langle \partial^{l_1}_i \psi(\boldsymbol{\theta}) | \partial^{l_2}_j \psi(\boldsymbol{\theta}) \rangle \Big].
    \label{eq:1TermApp}
\end{equation}

When \(l_1 = l_2\), this term can be efficiently estimated on a quantum computer. Specifically, for a given layer \(l\), all terms for every \(i\) and \(j\), along with those in Eq. \ref{eq:2RHS}, can be computed simultaneously using a single quantum circuit. This approach, introduced in \cite{stokes2020quantum}, is known as the block-diagonal approximation of the QFIM. Consequently, if the quantum circuit consists of \(L\) different blocks, estimating the block-diagonal approximation of the QFIM would require \(L\) distinct quantum circuits, each with the same number of qubits as the unitary \(U(\boldsymbol{\theta})\) that prepares the parameterized quantum state.

Conversely, if \(l_1 \neq l_2\), we have demonstrated that for commuting-block circuits, Eq. \ref{eq:1TermApp} can be estimated using the protocol described in Sec. \ref{Protocol}. Furthermore, for a fixed \(l_1\) and \(l_2\), all terms for every \(i\) and \(j\) can be simultaneously estimated using the same quantum circuit. This circuit requires an additional auxiliary qubit compared to the circuit that prepares the parameterized quantum state. Thus, \(\frac{L(L-1)}{2}\) quantum circuits would be needed to estimate all the upper block-diagonal terms.

In the following, we illustrate a procedure for simultaneously estimating the expectation values of commuting observables on a quantum computer by diagonalizing them in a common basis. 

Assume we have a set of commuting observables \(\{O_j\}\) and we wish to measure their expectation values with respect to a quantum state \(|\psi\rangle\). Denoting by \(\{ |\lambda_i\rangle \}\) the common eigenbasis in which the \(O_j\) are diagonal, each operator can be expressed as:

\begin{equation}
    O_j = \sum_i \lambda_i(O_j) |\lambda_i\rangle \langle \lambda_i|,
    \label{eq:Op_Decomp}
\end{equation}
where \(\{ \lambda_i(O_j) \}\) are the corresponding eigenvalues of \(O_j\).

To estimate the expectation values of all observables simultaneously, we apply a unitary transformation that rotates \( |\psi\rangle \) into the eigenbasis \( \{ |\lambda_i\rangle \} \). This eigenbasis always exists in theory, but can be challenging to determine in practice. However, for a set of commuting Pauli products, one can efficiently identify unitary operators from the Clifford group that rotate the observables into their shared eigenbasis~\cite{Yen2020,Yen2023}. Measurements are then performed in this basis to estimate the corresponding probability distribution:

\begin{equation}
    P(i) = |\langle \lambda_i  |\psi\rangle|^2.
    \label{eq:Prob_Distr}
\end{equation}

Using Eq. \ref{eq:Op_Decomp}, the expectation value of each observable can be rewritten as:

\begin{equation}
    \Big \langle O_j\Big \rangle_{|\psi\rangle} = \mathbb{E}_{P(i)}[\lambda_i(O_j)].
    \label{eq:Expect_Val}
\end{equation}

Thus, estimating all expectation values of \(\{O_j\}\) in parallel requires a single quantum circuit that prepares the state \(|\psi\rangle\) and rotates it into the eigenbasis \(\{ |\lambda_i\rangle \}\). By estimating the probability distribution \(P(i)\), we can compute all expectation values using Eq. \ref{eq:Expect_Val}.

In practice, \(P(i)\) is estimated by performing \(M\) measurements, yielding a set of outcomes \(\{i_1, \dots, i_M\}\). The expectation value is then approximated as:

\begin{equation}
    \Big \langle O_j\Big \rangle_{|\psi\rangle} \approx \frac{1}{M} \sum_{k=1}^{M} \lambda_{i_k}(O_j).
\end{equation}

By the Central Limit Theorem, the variance of the sample mean of a bounded random variable scales as \(\mathcal{O}(1/M)\), meaning the accuracy of this approximation improves as more samples are collected. 

Therefore, to estimate the full QFIM, classical post-processing of data from \(\frac{L(L+1)}{2}\) distinct quantum circuits is required.  \hfill \(\square\)

\end{document}